# The effect of Nb and O on the martensitic transformation in Ti–Nb–O alloys

**Kristián Šalata,[a]* Dalibor Preisler,[a] Josef Stráský,[a] Jiří Kozlík,[a] Lukáš Horák[b] and Václav Holý[b]**

[a]Department of Physics of Materials, Charles University in Prague, Ke Karlovu 3, Prague, Czechia, and [b]Department of Condensed Matter Physics, Charles University, Ke Karlovu 5, Prague, Czechia. *Correspondence e-mail: kristian.salata@matfyz.cuni.cz

This study examines the influence of niobium and oxygen on phase stability, crystal structure and martensitic transformation pathways in Ti–Nb–O alloys. A series of Ti–(8–28)Nb–(0–3)O (at.%) alloys were prepared and solution-treated in the $\beta$-phase field. The microstructure and crystallography were characterized by X-ray diffraction (XRD), electron microscopy and reciprocal-space mapping. A 2D-XRD orientation simulation approach was applied to distinguish all 12 crystallographically equivalent $\alpha''$-martensitic variants originating from a single prior $\beta$ grain, enabling detailed diffraction analysis. This method further allowed quantitative evaluation of the atomic shuffle parameter $y$, describing the $\beta \rightarrow \alpha''$ transformation. The results demonstrate that Nb primarily governs $\alpha''$-martensite evolution. Increasing Nb stabilizes the $\beta$ phase and shifts the $\alpha''$ structure toward higher symmetry, as reflected by systematic changes in lattice parameters and increasing shuffle parameter $y$, indicating suppression of the transformation to the hexagonal $\alpha'$ phase. Oxygen, in contrast, modifies the transformation pathways. At lower Nb contents, it suppresses $\omega$-phase formation and promotes $\beta \rightarrow \alpha''$ transformation, while at higher Nb levels it inhibits long-range martensitic transformation, resulting in retained $\beta$ or competing $\omega$ phase. These effects are attributed to local lattice distortions induced by interstitial oxygen.

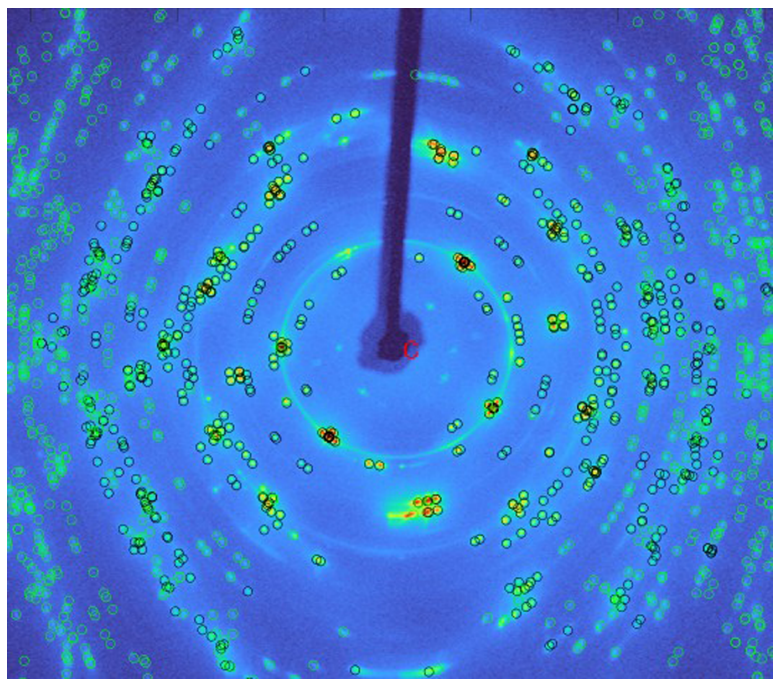

## 1. Introduction

Titanium alloys represent a versatile class of materials whose mechanical and functional properties are governed by a complex interplay of phase transformations. Depending on alloy composition, thermomechanical processing and interstitial solute content, these materials can exhibit a wide range of microstructures, from equilibrium phases ($\alpha$, $\beta$) to metastable and non-equilibrium states ($\alpha'$, $\alpha''$, $\omega$) (Lütjering & Williams, 2007). Such microstructural diversity enables tailoring of their mechanical properties for applications spanning aerospace (Boyer & Briggs, 2005), biomedical (Abdel-Hady Gepreel & Niinomi, 2013; Tane *et al.*, 2011) and shape-memory technologies (Costanza & Tata, 2020).

Conventional $\alpha$+$\beta$ titanium alloys – most prominently Ti–6Al–4V – remain widely used structural and biomedical materials because of their favourable mechanical performance. Nevertheless, their suitability for long-term implantation is debated due to compositional and mechanical limitations. Young's modulus for these alloys is of the order of 110–115 GPa, far exceeding that of human bone, which promotes stress shielding whereby the implant carries the majority of load and the surrounding tissue undergoes resorption and weakening (Bittredge *et al.*, 2022).

In addition, concerns have been raised regarding alloying elements: it has been reported that vanadium and aluminium ions released from Ti–6Al–4V may contribute to adverse biological responses, and the presence of these elements has been associated with cytotoxic or neurological effects (Guo *et al.*, 2015). These limitations have motivated extensive research into metastable $\beta$-type titanium alloys designed with biocompatible stabilizers such as Nb, Mo, Ta and Zr. Such alloy systems aim to reduce elastic stiffness while maintaining strength and ductility (Yang *et al.*, 2026; Preisler *et al.*, 2023; Stráský *et al.*, 2022).

Among these advanced alloys, the class known as 'gum metals' (*e.g.* Ti–23Nb–2Zr–0.7Ta–1.2O at.%) have attracted significant attention due to their exceptional mechanical properties, including low elastic modulus, high strength, superelasticity, excellent cold workability and Invar-like behaviour (Kuramoto *et al.*, 2006). Unlike conventional martensite, which forms via a diffusionless displacive mechanism involving a coupled shear and shuffle (Zheng *et al.*, 2022; Dong *et al.*, 2024), the presence of defects in such gum metals caused by increased oxygen content and/or $\beta$-stabilizer content can frustrate long-range ordering, leading to the formation of nanoscale martensitic domains or 'strain-glass' states instead of macroscopic martensite plates, as reported in several studies (Zheng *et al.*, 2016*b*; Liang *et al.*, 2019; Wang *et al.*, 2014).

Interstitial oxygen plays a nontrivial role in the phase stability and transformation behaviour of metastable $\beta$-titanium alloys. While it is traditionally categorized as an $\alpha$-stabilizing element (Lütjering & Williams, 2007), experimental and theoretical studies demonstrate that its influence is more nuanced in Nb-containing systems in terms of $\beta \rightarrow \alpha''$ transformation (Kawano *et al.*, 2019).

Although several types of interstitial sites exist, the most stable are octahedral sites for both hexagonal close-packed (h.c.p.) $\alpha$ and body-centred cubic (b.c.c.) $\beta$ phases (Ouyang *et al.*, 2018). Oxygen atoms occupying interstitial sites generate local lattice distortions and strain fields that can stabilize orthorhombic $\alpha''$-martensite by modifying its lattice parameters and increasing the reverse transformation temperature (Tahara *et al.*, 2016). At the same time, these strain fields promote the formation of nanoscale modulated domains in the parent $\beta$ phase, which act as barriers to the development of long-range martensitic transformation (Tahara *et al.*, 2011). Such effects go beyond classical solid-solution strengthening, reflecting a coupling between local structural distortion, phase-transformation kinetics and deformation mechanisms (Chong *et al.*, 2023; Wang *et al.*, 2021).

Despite considerable progress in the development of metastable $\beta$-Ti systems, several aspects of phase stability and transformation behaviour remain unresolved. The precise relationship between $\beta$-stabilizer content, oxygen concentration and $\alpha''$-martensite formation has still not been fully explored. Additionally, the competition between $\alpha''$-martensite and athermal $\omega$-phase formation – a common issue in lean $\beta$ alloys – requires further investigation, as $\omega$-phase formation can embrittle the material and degrade functional properties. Addressing these challenges necessitates a systematic study of lattice parameter evolution, phase stability and phase formation across a range of compositions.

In this context, the present work provides a comprehensive investigation of Ti–Nb–O alloys, aimed at clarifying the relationship between alloy chemistry, crystal structure and martensitic microstructure. By combining diffraction-based phase identification, determination of the crystal structures of the phases and microstructural observations, the study seeks to investigate the distinct roles of niobium and oxygen and their specific impact on the crystal structure. Special focus is placed on the interplay between these elements in governing the martensitic transformation and on the competitive formation of $\alpha''$, $\beta$ and $\omega$ phases. By uncovering the underlying mechanisms of lattice stability in the presence of interstitial solutes, this work contributes to the broader understanding required for the rational design of next-generation low-modulus biomedical titanium alloys.

## 2. Crystallographic relation between $\beta$ and $\alpha''/\alpha'$ phase

The $\beta \rightarrow \alpha'/\alpha''$ martensitic transformation in metastable b.c.c. alloys is most rigorously interpreted as a diffusionless co-operative lattice instability governed by coupled shear–shuffle modes. Crystallographically, the transformation pathway can be described as the superposition of two distinct kinematic components. The first corresponds to a lattice-invariant shear occurring approximately along the $\{112\}\langle 111\rangle_{\beta}$ system (Burgers, 1934), which accommodates the major shape deformation and establishes the macroscopic crystallographic correspondence between parent and product phases. The second component consists of a short-range atomic shuffle involving displacements within $\{110\}_{\beta}$ planes along $\langle 1\overline{1}0\rangle_{\beta}$-type directions (Lütjering & Williams, 2007; Wayman, 1964). These modes are required to satisfy the basic rearrangement necessary for the formation of a close-packed stacking sequence. The primary distinction between martensitic products lies in the extent of this displacement: while a complete shear/shuffle (where atoms reach the ideal hexagonal sites) results in the formation of the hexagonal $\alpha'$ phase, orthorhombic $\alpha''$-martensite is characterized by a lower degree of transformation relative to this hexagonal limit. Therefore, orthorhombic martensite ($\alpha''$) has a lower symmetry than either the $\beta$ (b.c.c.) or $\alpha'$ (h.c.p.) phases.

This feature can be understood by considering the subgroups of the b.c.c. $\beta$ phase and the h.c.p. $\alpha'$ phase. Their space groups are $Im\overline{3}m$ ($\beta$) and $P6_3/mmc$ ($\alpha'$) and, according to the literature (Bendersky *et al.*, 1994), the intersection subgroup is the orthorhombic *Cmcm* with its *c* axis parallel to one of the $\langle 110\rangle_{\beta}$ directions. It can be visualized as the structure shown in Fig. 1, which is close to h.c.p. but differs in its symmetry and the relative positions of the atoms in the basal plane.

From the symmetry point of view, the structure corresponding to space group *Cmcm* ($\alpha''$) can transform into a structure with a higher symmetry by a change in lattice parameters and an alternating shuffle of the opposite $\{110\}_{\beta}$ planes. This transition, primarily (but not exclusively) made up of $\{110\}_\beta\langle 1\bar{1}0\rangle_\beta$ shuffle, can be visualized by drawing the corresponding *Cmcm* unit cell onto the $(110)_\beta$ and $(0001)_{\alpha'}$ planes (Fig. 1). The degree of shuffling is reflected in the internal $y$ coordinate of the Wyckoff position 4*c*: $(0, y, \frac{1}{4})$ (Hahn, 2005). Using this correspondence, the $\beta$ phase (b.c.c.) could be characterized by $y = \frac{1}{4}$ [Fig. 1(*a*)] and $\alpha'$ (h.c.p.) by $y = \frac{1}{6}$.

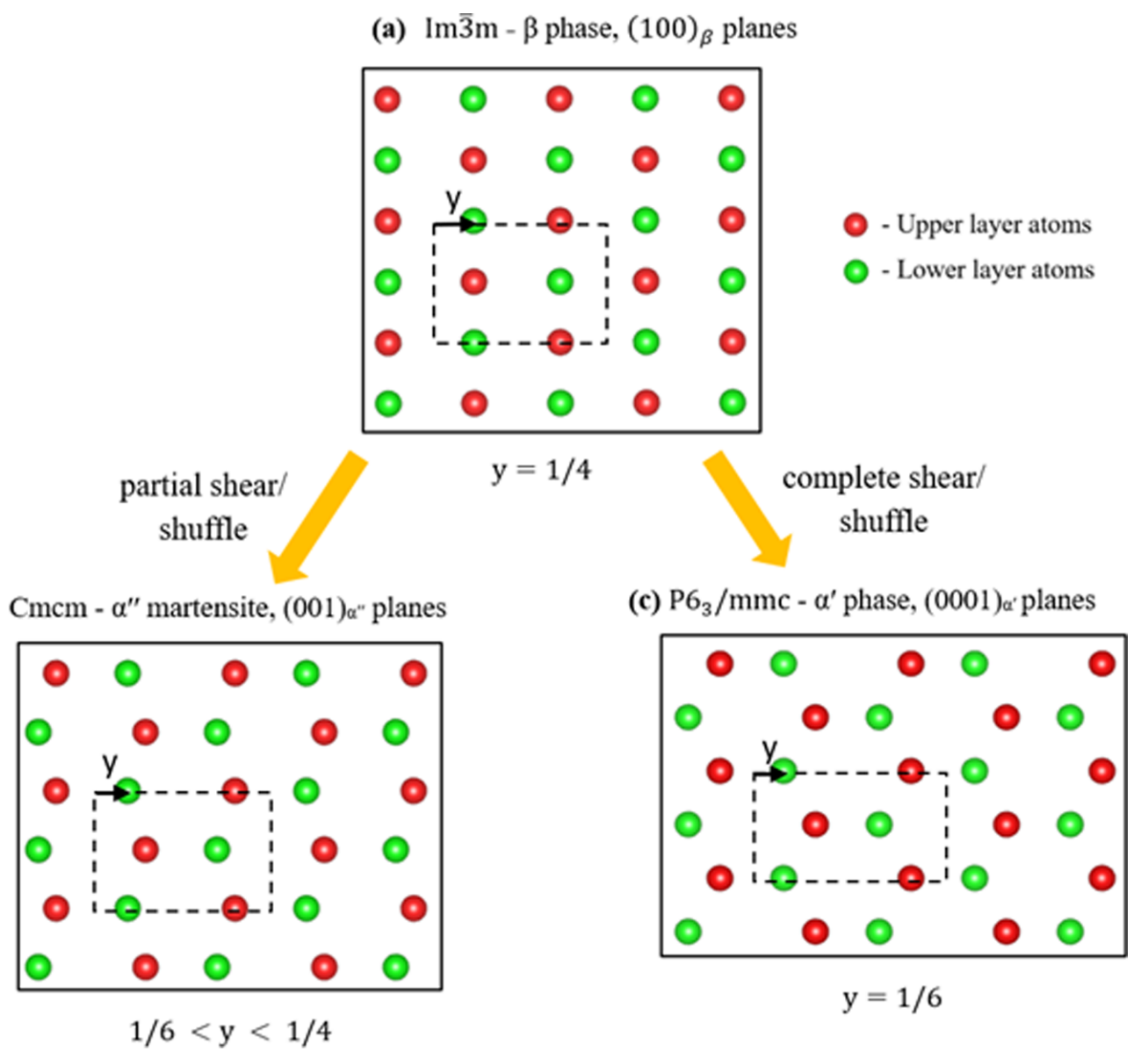

**Figure 1**
Correspondence between the parent $\beta$ phase and orthorhombic $\alpha''$- or h.c.p. $\alpha'$-martensite. Dashed rectangles represent the corresponding unit cell of the *Cmcm* space group.

At $y = \frac{1}{6}$ the shuffle is complete and the structure degenerates into the hexagonal $\alpha'$ phase (provided the lattice ratio also matches – $b/a$ must be equal to $\sqrt{3}$ to have a 60° angle between three neighbouring atoms in one plane and thus hexagonal symmetry).

At $y = \frac{1}{4}$ the atoms remain in their relative positions from the parent $\beta$ phase, representing a state where the shuffle has not yet occurred.

To evaluate the progression of the transformation in our Ti–Nb alloys, we utilize the orthorhombicity parameter to determine the deviation from the hexagonal limit:

$$\eta_S = \frac{y - 1/4}{1/6 - 1/4} = 3 - 12y.$$

This parameter normalizes the shuffle magnitude, where $\eta_S = 0$ corresponds to the complete shuffle of the hexagonal $\alpha'$ phase and $\eta_S = 1$ corresponds to the zero-shuffle state of the b.c.c. $\beta$ phase.

Whether $\alpha'$ or $\alpha''$ will appear upon quenching depends mainly on the concentration of solute atoms. In the Ti–Nb binary system, the compositional boundary between $\alpha'$ and $\alpha''$ martensitic phases has been reported from first-principles calculations to lie at approximately 5.7 at.% Nb (Pathak *et al.*, 2014), which implies that Nb-lean alloys with lower concentrations will transform to hexagonal martensite ($\alpha'$ phase) and Nb-rich alloys to orthorhombic $\alpha''$. The orientation relation between $\beta$ and $\alpha''$ can be described as

$$(110)_\beta \parallel (001)_{\alpha''}, \quad [1\bar{1}1]_\beta \parallel [1\bar{1}0]_{\alpha''}.$$

This symmetry reduction, governed by the loss of specific symmetry elements, results in 12 crystallographically equivalent variants of $\alpha''$.

**Table 1**
Results from the measurement of oxygen and nitrogen content in all samples.

Data are presented as average values with standard deviations.

| Sample | O (at.%) | $\sigma_O$ (at.%) | N (at.%) | $\sigma_N$ (at.%) |
|---|---|---|---|---|
| Ti–8Nb | 0.21 | 0.04 | 0.06 | 0.05 |
| Ti–12Nb | 0.18 | 0.01 | 0.08 | 0.10 |
| Ti–12Nb–1O | 0.91 | 0.05 | 0.05 | 0.02 |
| Ti–12Nb–2O | 1.65 | 0.11 | 0.05 | 0.01 |
| Ti–12Nb–3O | 2.75 | 0.01 | 0.06 | 0.01 |
| Ti–16Nb | 0.18 | 0.03 | 0.02 | 0.01 |
| Ti–16Nb–1O | 0.89 | 0.03 | 0.08 | 0.06 |
| Ti–16Nb–2O | 1.81 | 0.16 | 0.03 | 0.03 |
| Ti–16Nb–3O | 2.64 | 0.03 | 0.04 | 0.01 |
| Ti–20Nb | 0.21 | 0.05 | 0.02 | 0.01 |
| Ti–20Nb–1O | 0.92 | 0.04 | 0.02 | 0.01 |
| Ti–20Nb–2O | 1.83 | 0.48 | 0.05 | 0.01 |
| Ti–20Nb–3O | 2.71 | 0.09 | 0.06 | 0.04 |
| Ti–24Nb | 0.23 | 0.11 | 0.03 | 0.02 |
| Ti–28Nb | 0.22 | 0.11 | 0.04 | 0.02 |

## 3. Methods

Binary and ternary Ti–$x$Nb–$y$O alloys with compositions ranging from 8 to 24 at.% of Nb and 0 to 3 at.% of O were prepared by a conventional arc melting method from 99.9% pure Ti pellets, Ti–45 Nb pre-alloy and $TiO_2$ powder in a Zr-gettered Ar atmosphere. All ingots were remelted 6–10 times to ensure complete dissolution of Nb and O within the melt. A cold copper crucible was used to produce fifteen 8 g buttons; each button was homogenized at 1200°C in the single $\beta$-phase domain under a vacuum ($5 \times 10^{-5}$ Pa) for 12 h and subsequently water quenched. The specimens were prepared to obtain large single $\beta$ grains, which ensures that the diffraction signal originates from a single crystallographic domain. This allows individual martensitic variants to be distinguished and their orientation to be resolved, providing essential information for accurate crystal structure refinement. Analyses of the O content were performed using a carrier gas hot extraction method on three samples of each alloy; the results are summarized in Table 1.

For scanning electron microscopy (SEM) and light microscopy observations, the specimens were prepared following standard metallographic techniques, using SiC grinding papers with grit size up to 4000, followed by a final polish on a metallographic polishing cloth with OP-S suspension containing approximately 20% $H_2O_2$ for 15 min. All samples were analysed on an FEI QUANTA scanning electron microscope equipped with a field emission gun at 15 kV or a Zeiss reflective polarized light microscope.

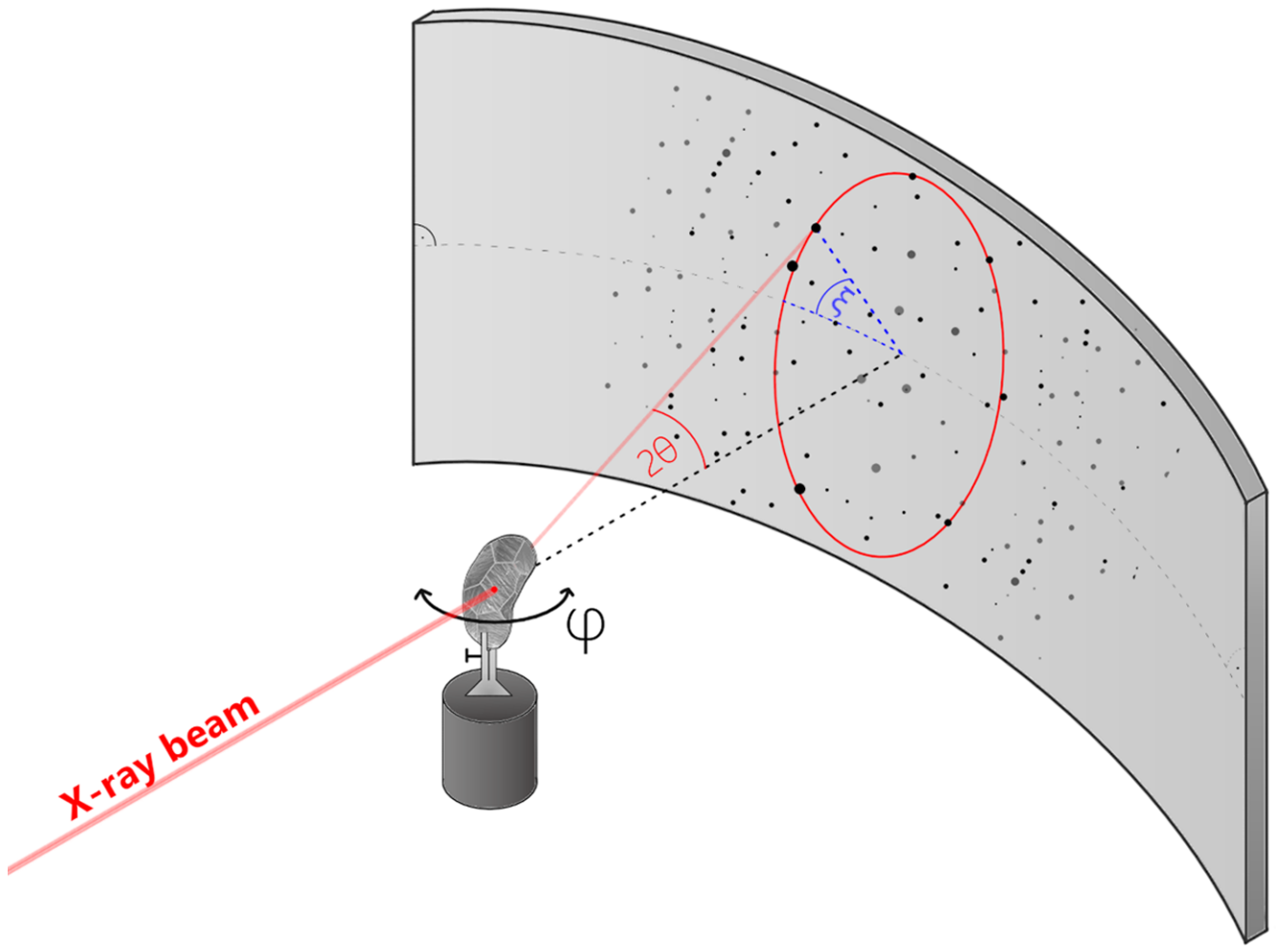

**Figure 2**
Design of the experiment for crystal structure determination with a 2D detector image plate.

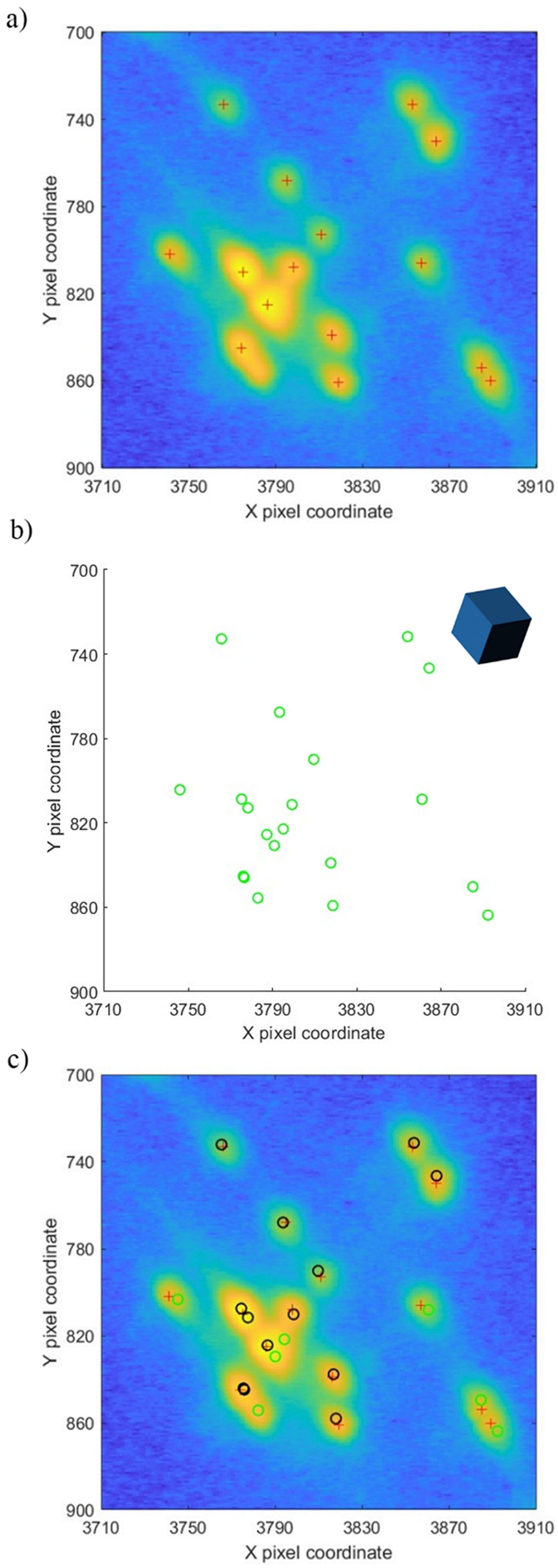

**Figure 3**
(*a*) Part of an image plate showing identified centres of individual peaks used for structure orientation determination. (*b*) Reciprocal-space map simulation for different orientations of the parent $\beta$ grain. (*c*) Final reciprocal-space map simulation, with black circles indicating simulation–experiment pair matches and green circles denoting predictions lacking the pair within a certain distance.

To determine the crystal structure parameters, an X-ray diffractometer (Rigaku R-axis RAPID II) was used, equipped with an Mo anode and a graphite monochromator operated at 50 kV and 40 mA. Diffraction patterns were recorded on the cylindrical 2D image plate with the arrangement presented in Fig. 2.

### 3.1. Orientation analysis

Crystal structure determination conventionally relies on the analysis of a single crystal with a well defined chemical composition, followed by the extraction of integrated diffraction intensities and calculation of structure factors. These structure factors are subsequently used to determine atomic positions and to refine the crystallographic model iteratively. However, the metastable character of martensitic phases generally precludes the growth of suitable martensitic single crystals. As a result, conventional single-crystal diffraction approaches cannot be readily applied, significantly complicating the determination of martensitic crystal structures.

Owing to these constraints, an alternative experimental strategy analogous to single-crystal X-ray diffraction (XRD) was employed. A parent prior $\beta$ grain with arbitrary orientation which fully transformed into $\alpha''$-martensite was selected. All $\alpha''$-martensite laths preserved a well defined crystallographic relationship with the parent $\beta$ orientation. As the martensitic variants are mutually related through the orientation relationship, the presence of retained $\beta$ phase was not required for determination of the orientation. This approach enabled us to distinguish the individual martensitic variants and allowed reciprocal-space maps corresponding to each variant to be refined separately.

Comparison of the simulated diffraction patterns with the experimentally measured $\varphi$-integrated 2D reciprocal-space map, acquired while the sample was rotated over a range of $-30^\circ$ to $30^\circ$, enabled indexing of all observed reflections and assignment of the corresponding rotation angles at which they were recorded. This information was essential for determining the absorption correction factor, a critical parameter in the subsequent analysis.

Our simulation procedure can be summarized in the following steps:

(i) *Peak identification and selection.* The angular coordinates $(\theta, \xi)$ of all non-overlapping reflections were determined from the $x$–$y$ coordinates of the image plate. Only reflections sufficiently separated from neighbouring peaks were considered to ensure accurate measurement [see Fig. 3(*a*)].

(ii) *Reciprocal-space map simulation.* The experimental peaks positions were simulated considering all 12 crystallographically related martensitic variants, accounting for different possible starting orientations of the parent $\beta$ grain [see Fig. 3(*b*)].

(iii) *Simulation–measurement matching.* The number of matched reflection pairs between simulation and experiment was calculated for each $\beta$ orientation as a scoring metric. Black circles indicate matched peaks; green circles denote simulation peaks lacking a pair within a certain distance [Fig. 3(*c*)].

To avoid nonlinear scaling of rotation in spherical coordinates, we employed the *Icosahedron Mesh* function in MATLAB, which provided a more uniform step size for structure orientation. The coordinates of all the reciprocal vectors for the $\alpha''$ phase based on lattice parameters were derived from integrated 1D profiles over a $(0, 2\pi)$ $\xi$ interval at a constant $2\theta$ angle.

Using crystallographic orientation relationships between the $\beta$ and $\alpha''$ phases, we predicted and simulated diffraction patterns for all 12 mutually oriented martensitic variants, with precise determination of $2\theta$ and $\xi$ angles, simulating the experimental $\varphi$ rotation from $-30^\circ$ to $30^\circ$ for each reciprocal point.

### 3.2. Crystal structure determination

The preceding orientation analysis provided the necessary crystallographic framework to isolate the individual $\alpha''$ variants. Having successfully mapped the experimental peaks to the corresponding diffraction of the theoretical lattice orientation, the next objective was the quantitative determination of the internal atomic positions, specifically the $y$ coordinate of the atomic Wyckoff position 4$c$.

To determine the atomic positions, it is necessary to extract the experimental (observed) structure factor $F_{hkl}^{\rm obs}$ for each identified peak. The structure factor is defined as

$$F_{hkl} = \sum_{n=1}^{N} f_n \exp\left[2\pi i(hx_n + ky_n + lz_n)\right],$$

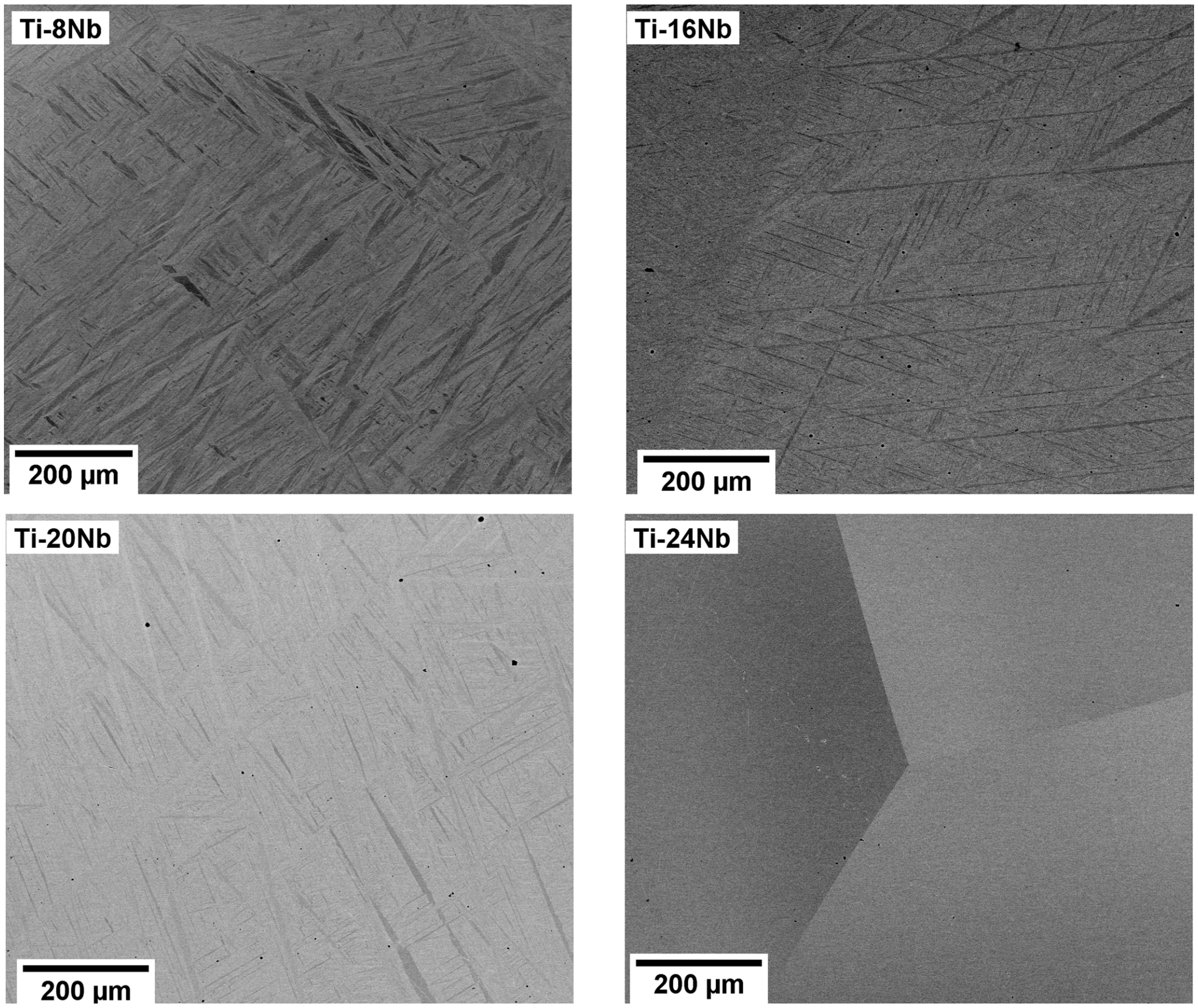

**Figure 4**
BSE images of Ti–$x$Nb samples without added oxygen after water quenching from 1200°C.

where $f_n$ is the atomic scattering factor of atom $n$, $hkl$ denote the Miller indices of the reflecting lattice plane and $(x_n, y_n, z_n)$ are fractional atomic coordinates within the unit cell. In the *Cmcm* space group, these coordinates are specifically defined by the $(0, y, \frac{1}{4})$ $(0, -y, \frac{3}{4})$ $(\frac{1}{2}, \frac{1}{2}+y, \frac{1}{4})$ $(\frac{1}{2}, \frac{1}{2}-y, \frac{3}{4})$ positions (Hahn, 2005). This extraction process relies on the measured integrated intensities $I_{\rm measured}$, the determination of which is described in detail in Section 4.4. Within the framework of the classical kinematic theory of XRD, the structure factor was then modelled using these integrated intensities and the following equation:

$$I_{\rm measured} = I_0 \, A \left|F_{hkl}\right|^2 Lp,$$

where $A$ stands for the absorption correction factor, $L$ is the Lorentz factor, $p$ is the polarization factor and $I_0$ is an overall scale factor.

## 4. Results and discussion

### 4.1. Variation of Nb

Qualitative phase analysis was carried out on the X-ray diffraction data. The 2D image plate patterns were integrated over a constant diffraction angle ($2\theta$), providing a sensitive probe of the structural constituents. Complementary SEM observations were employed to correlate the identified phases with their corresponding morphological features.

The martensitic structure observed by SEM (Fig. 4) appears to be characteristic of 'massive'-type martensite consisting of lamellar colonies containing several parallel plates. XRD results (Fig. 5) confirm the presence of athermal $\omega$ ($\omega_{\rm ath}$) alongside $\alpha''$-martensite across the Ti–(8–16)Nb compositional range. A significant change in the lattice parameters shifts the individual peaks from their position presented at the bottom, almost merging the structure into the h.c.p. $\alpha'$ structure at the Ti–8Nb composition [but still not fully, as two peaks between (27°, 28°) indicate]. The observed peak at $2\theta \simeq 25°$ provides a definite qualitative marker for the presence of the $\omega_{\rm ath}$ phase. Owing to the nanometric size and low volume fraction of $\omega_{\rm ath}$ precipitates, the weaker $\omega$ reflections are largely suppressed and only the most intense peaks remain detectable.

This behaviour can be rationalized in terms of $\beta$-phase stability and martensitic transformation thermodynamics. In the Ti–8Nb alloy, the relatively high martensite start temperature ($M_{\rm S}$) promotes nearly complete $\beta \rightarrow \alpha''$ transformation during quenching, thereby limiting the amount of retained parent $\beta$ phase available for $\omega_{\rm ath}$ formation (Kozlík *et al.*, 2021). In contrast, the $\beta$ phase in Ti–16Nb is more strongly stabilized by the higher Nb content, which suppresses the lattice instability and $\{111\}_\beta$ plane collapse associated with $\omega_{\rm ath}$ formation. Consequently, the $\omega_{\rm ath}$ volume fraction is reduced at high Nb concentrations. The intermediate Ti–12Nb composition thus represents a specific equilibrium in the martensitic suppression at low stability of the parent $\beta$ phase, leading to enhanced $\omega_{\rm ath}$ formation.

In the present work, $\omega_{\rm ath}$ is clearly detected up to Ti–16Nb and no distinct $\omega$ reflections were observed in Ti–20Nb (Fig. 6). Notably, $\alpha''$-martensite is still present at this composition.

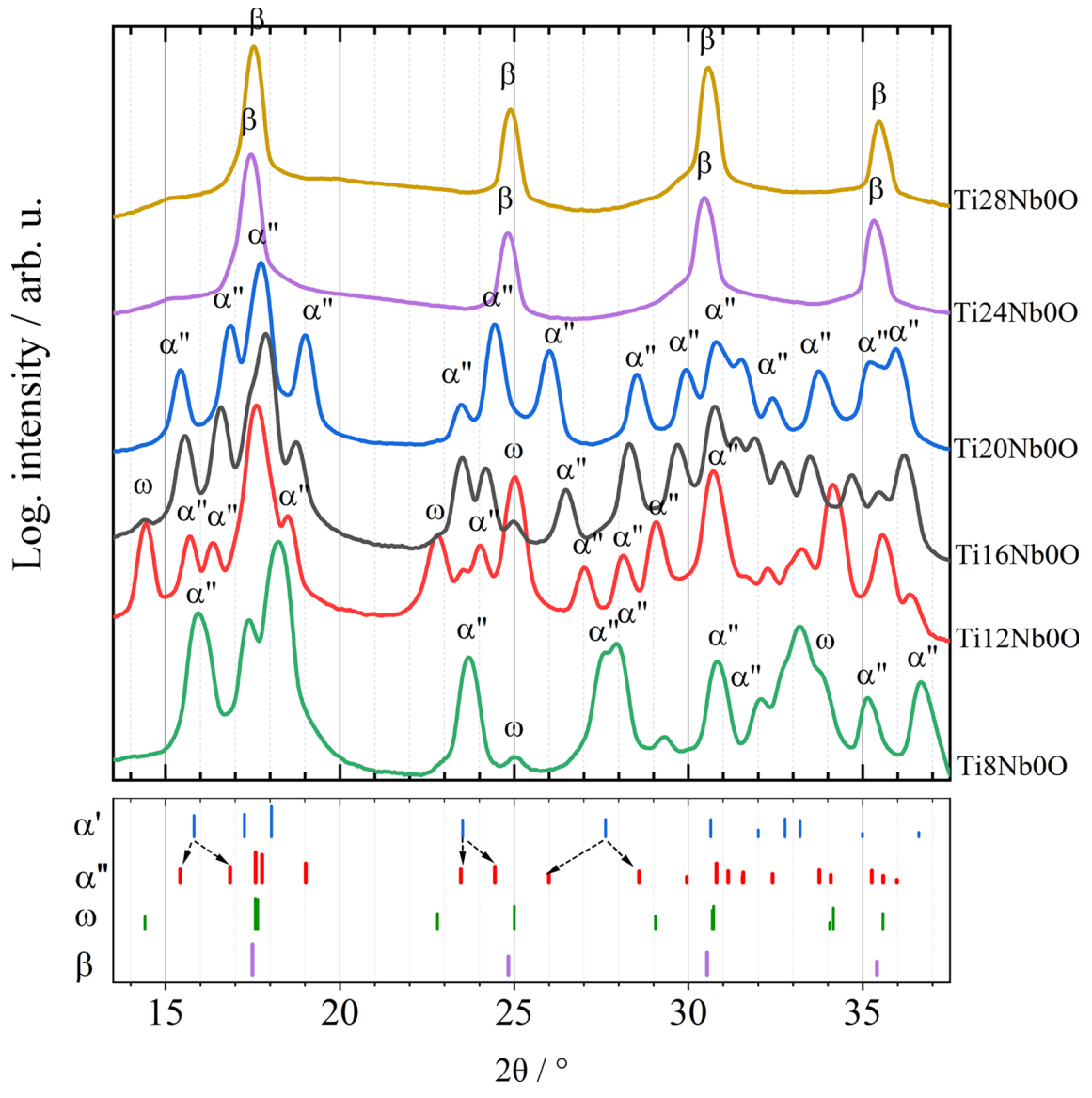

**Figure 5**
XRD integrated profile over constant $2\theta$ angle of Ti–$x$Nb alloys after quenching from 1200°C. The legend on the right-hand side indicates which curve belongs to which sample. Black arrows show the splitting of several peaks due to the lower symmetry of an orthorhombic structure ($\alpha''$) over h.c.p. ($\alpha'$).

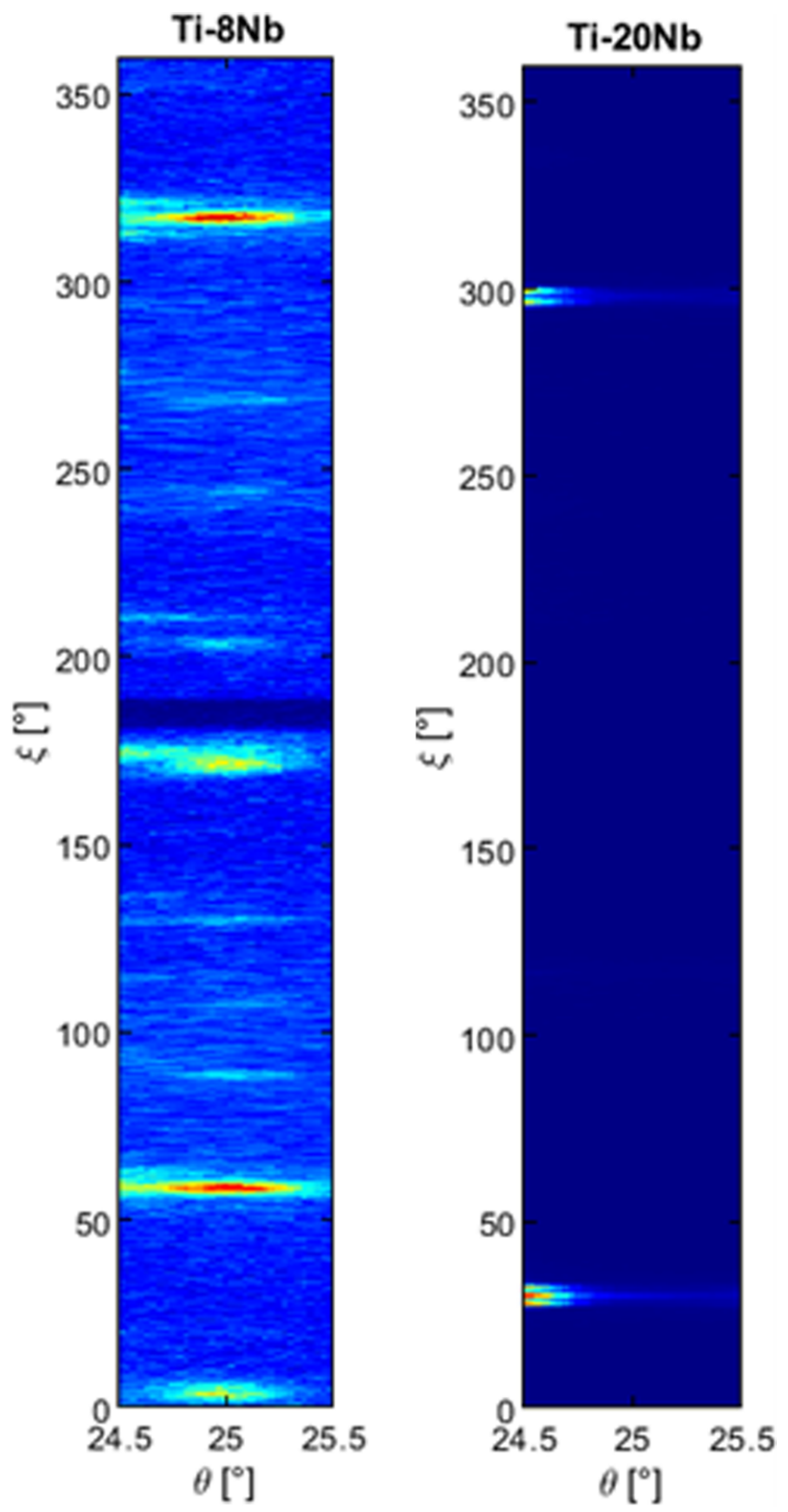

**Figure 6**
Comparison of diffraction at $2\theta = 25°$ from Ti–8Nb (left; $\omega$ peaks present) and Ti–20Nb (right; no $\omega$ peaks observed). Axes correspond to $2\theta/\xi$ angles from Fig. 2.

**Table 2**
Lattice parameters of the $\alpha''$ phase in Ti–Nb samples determined from an integrated profile obtained by XRD.

| Nb (at.%) | $a$ (Å) | $b$ (Å) | $c$ (Å) | $b/a$ |
|---|---|---|---|---|
| 8 | 2.964 | 5.068 | 4.687 | 1.710 |
| 12 | 3.047 | 4.999 | 4.679 | 1.641 |
| 16 | 3.103 | 4.935 | 4.663 | 1.590 |
| 20 | 3.160 | 4.850 | 4.650 | 1.535 |

Previous studies have shown that increasing Nb content progressively suppresses the $\beta \to \alpha''$ martensitic transformation while still allowing $\omega_{ath}$ to form within the retained $\beta$ matrix in a certain compositional window (Lyon *et al.*, 1983). Conversely, our results may indicate sufficient stabilization to suppress the $\omega$ phase, but not enough to supress the $\beta \to \alpha''$ transformation, most probably depending on alloy composition including interstitials. Diffraction patterns from Ti–24Nb and Ti–28Nb indicate solely the $\beta$ phase, consistent with SEM observations.

To quantify the influence of Nb on orthorhombic martensite, lattice parameters were determined via Rietveld refinement (*X'Pert HighScore Plus*; Degen *et al.*, 2014) of integrated profiles treated as powder data; the results are summarized in Table 2. Fig. 7 shows the dependence of the lattice parameters on Nb content. Since lattice parameters depend primarily on peak positions, the results are considered reliable even without matching intensities. As shown in Fig. 7, parameter $a$ increases with Nb content, while $b$ and $c$ decrease. This behaviour, consistent with the literature (Brown *et al.*, 1964; Banumathy *et al.*, 2009; Bönisch *et al.*, 2014), cannot be explained by the atomic size differences between Ti and Nb. Instead, it reflects Nb's tendency to retain eightfold coordination characteristic of the $\beta$ phase (Brown *et al.*, 1964). Increasing Nb content therefore promotes orthorhombic symmetry $\alpha''$ in competition with the fully sheared h.c.p. $\alpha'$-martensite as it requires lower principal strains.

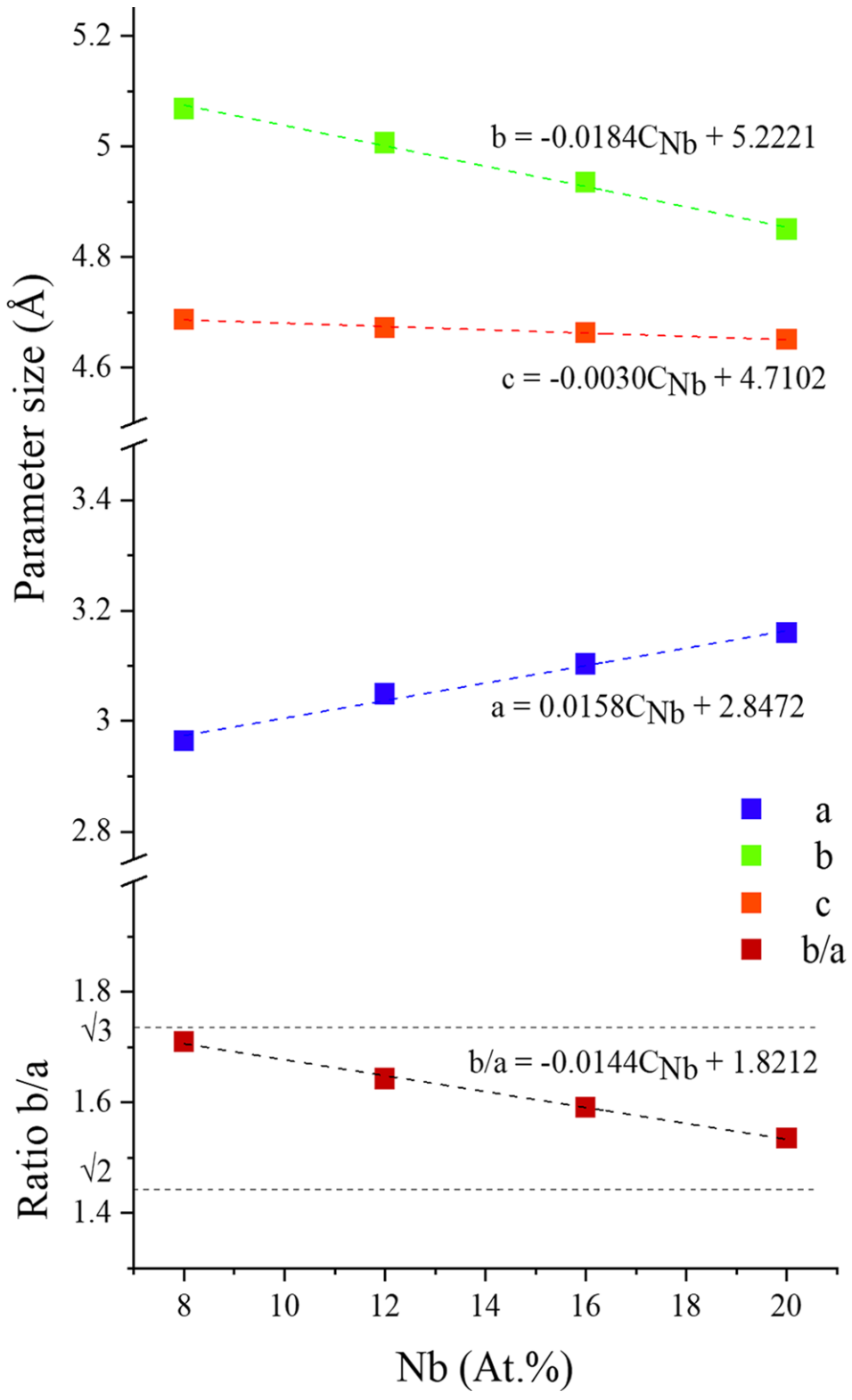

**Figure 7**
Change in the lattice parameters of martensite due to increasing content of Nb. The ratio $b'/a'$ is dimensionless. Coloured lines represent a linear fit.

The increase in $a$ and decrease in $b$ lattice parameters can be geometrically explained by $\{11\overline{2}\}\langle 111\rangle$ shear. This shear can be visualized via some of the $\{110\}_\beta$ planes. Fig. 8 schematically shows the partial shear in the $\langle 111\rangle$ direction together with above-mentioned $(11\overline{2})_\beta$ plane.

The higher the $b'/a'$ ratio is, the closer to a hexagonal symmetry structure we are. We can see that the axial ratio $b'/a'$ decreases with increasing Nb content, from 1.71 at 8 at.% of Nb to 1.53 at 20 at.% of Nb. Because hexagonal martensite has a $b'/a'$ ratio equal to $\sqrt{3} \simeq 1.73$ and the b.c.c. $\beta$ phase $\sqrt{2} \simeq 1.41$, the $\{11\overline{2}\}\langle 111\rangle$ shear is more suppressed with increasing Nb content.

The compositional dependence of the lattice parameter ratio can be described by the fitted relation $b/a = -0.0144C_{Nb} + 1.8212$, where $C_{Nb}$ is the Nb content in at.%. The line $b/a = \sqrt{3}$ represents the highest theoretical value of this ratio where the lattice virtually transforms into an $\alpha'$ lattice – the hexagonal limit. Taking the geometric condition corresponding to the b.c.c. limit $b/a = \sqrt{2}$, the critical concentration can be estimated to yield $C_{Nb}(\alpha'' \to \beta) \simeq 28$ at.% Nb. This extrapolated value is higher than the experimentally observed formation of a fully $\beta$ phase between 20 and 24 at.% of Nb. The discrepancy indicates that the linear compositional dependence described by the fitted equation is unlikely to remain valid once the alloy approaches the $\alpha''/\beta$ transition region, which has been reported previously (Bönisch *et al.*, 2014; Thoemmes *et al.*, 2021).

Representative backscattered electron (BSE)-SEM micrographs of $\beta$ solution-treated and water-quenched samples further contextualize the diffraction findings (Fig. 4). Alloys containing up to ~20 at.% Nb exhibited irregular martensitic plate morphologies characteristic of diffusionless transformation products, whereas higher-Nb alloys stabilized in the $\beta$ phase displayed coarse prior $\beta$ grains exceeding ~1 mm in size (lower magnification, not shown), consistent with prolonged high-temperature solution treatment. Such grain dimensions imply that the diffraction beam may effectively probe a single parent $\beta$ grain when the beam footprint is sufficiently confined, which is the case for the current study.

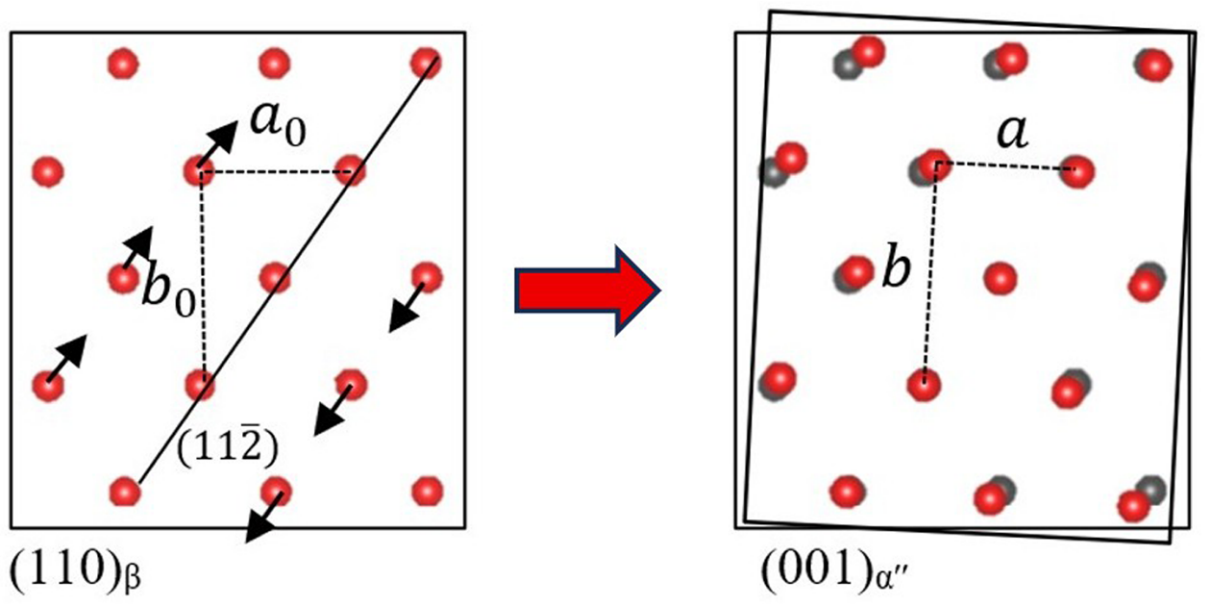

**Figure 8**
Transformation from the $\beta$ to the $\alpha''$ phase viewed from the $(110)_\beta$ plane composed of shear (black arrows) and shuffle (Zheng *et al.*, 2022).

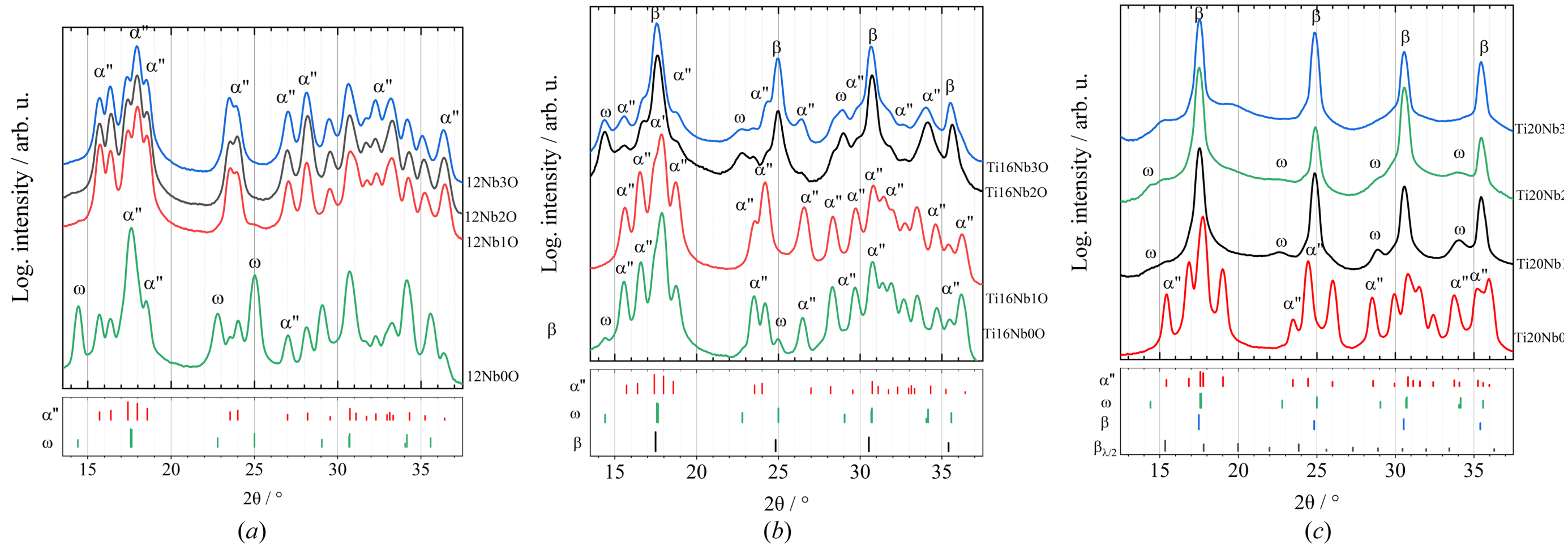

**Figure 9**
Integrated XRD diffraction patterns after $\beta$ solution heat treatment and quenching, (*a*) Ti–12Nb–$x$O, (*b*) Ti–16Nb–$x$O and (*c*) Ti–20Nb–$x$O.

### 4.2. Variation of O

Similar X-ray diffraction measurements were employed to resolve the phase constitution of Ti–Nb–O alloys, including minor-volume-fraction phases such as $\omega_{\rm ath}$ precipitates or finely scaled martensitic $\alpha''$ laths that are difficult to quantify by imaging alone.

For Ti–12Nb–$x$O alloys [Fig. 9(*a*)], $\omega$ reflections were detected only in the oxygen-free material and weakly in Ti–12Nb–1O, where a small peak appeared at $2\theta \simeq 25°$. No $\omega$ peaks were observed for higher oxygen contents. In contrast, diffraction signatures of martensitic $\alpha''$, manifested as multiple reflections, were present across all the compositions. Light microscopy images (Fig. 10) confirmed the martensitic morphology, supporting the XRD-derived phase identification. Together, these results demonstrate that oxygen has enhanced the $\beta{\rightarrow}\alpha''$ transformation pathway as opposed to the $\beta{\rightarrow}\omega$. This result is in agreement with a recent study (Tahara *et al.*, 2016), where oxygen addition led to $\alpha''$ stabilization once the macroscopic $\alpha''$ phase is formed.

A more complex evolution was observed in Ti–16Nb–$x$O alloys [Fig. 9(*b*)]. In the oxygen-free condition, the $\omega$ phase was present, whereas addition of 1 at.% O promoted full transformation of the $\beta$ phase into $\alpha''$-martensite in the same manner as in Ti–12Nb–$x$O. However, the Ti–16Nb–2O sample retained the $\beta$ phase after quenching, indicating partial inhibition of the $\beta{\rightarrow}\alpha''$ transformation; simultaneously, $\omega$ reflections re-emerged (visible peak near $2\theta \simeq 14.5°$). This behaviour suggests that increased oxygen content supresses martensite formation and redirects the transformation pathway toward $\beta{\rightarrow}\omega$ decomposition in regions where $\alpha''$ cannot form in a more stabilized $\beta$ alloy. A BSE-SEM micrograph of Ti–16Nb–2O (Fig. 11) revealed a mixed $\beta$+$\alpha''$ microstructure with localized martensitic laths, which is in agreement with diffraction results. A slight reduction in $\omega$ peak intensity between 2 and 3 at.% O may indicate renewed suppression at higher interstitial content, although quantitative verification would require more reliable peak intensity analysis and additional datasets.

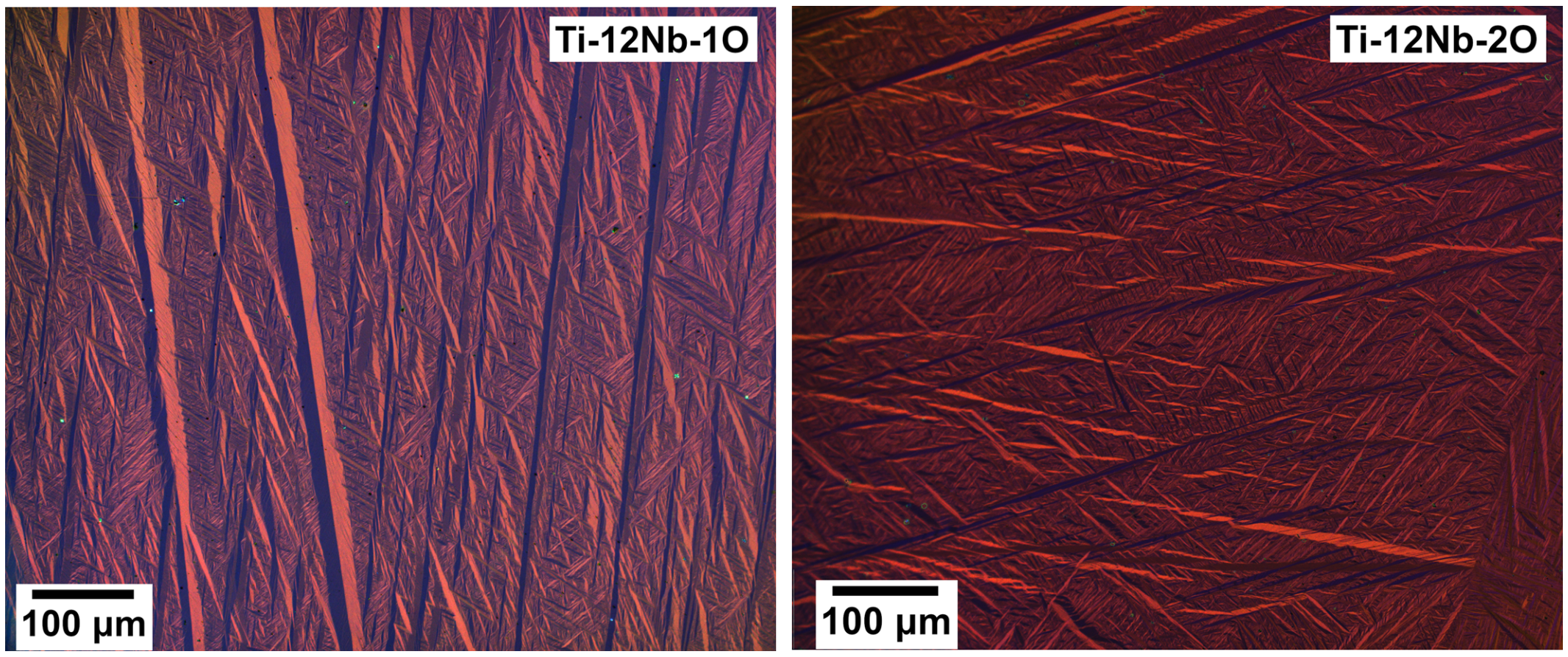

**Figure 10**
Polarized light microscopy images of Ti–12Nb–$x$O samples after solution treatment and water quenching.

These results are consistent with recent studies demonstrating that oxygen at higher concentrations suppresses the formation of macroscopic $\alpha''$-martensite upon quenching and instead promotes the formation of a nanodomain structure, visible by transmission electron microscopy (Tahara *et al.*, 2011; Zheng *et al.*, 2016*a*; Zheng *et al.*, 2018). This effect originates from oxygen atoms occupying octahedral interstitial sites in the parent $\beta$ phase, generating local anisotropic strain fields along the $\langle 100 \rangle_\beta$ directions, where there are three such directions. Because these interstitial sites are randomly and uniformly distributed, the associated strain fields are likewise random. This process results in the formation of nanoscale $\alpha''$-like domains with multiple crystallographic variants. The random distribution of these nanodomains prevents their cooperative growth into long-range martensite, thereby stabilizing the $\beta$ matrix. However, once macroscopic $\alpha''$ is formed, oxygen atoms further contribute to its stabilization by occupying octahedral sites in the distorted lattice (Tahara *et al.*, 2016).

Details of the phase constitution in higher-Nb alloys were similarly clarified for Ti–20Nb–*x*O [Fig. 9(*c*)]. In oxygen-free Ti–20Nb, no $\omega$ phase was detected, indicating that at elevated Nb content the $\omega$ transformation is already suppressed and the $\beta$ phase transforms directly to $\alpha''$ upon quenching. At approximately 1 at.% O, the martensitic transformation is suppressed and the diffraction patterns indicate predominantly retained $\beta$ with minor $\omega$ precipitation, consistent with the results presented by Tahara *et al.* (2016). Upon further increase in the oxygen content to 2 and 3 at.%, the $\omega$ phase disappears completely. This systematic disappearance demonstrates that oxygen effectively suppresses $\omega$ formation within the $\beta$-rich matrix, which was also reported by Tahara *et al.* (2016) and Paton & Williams (1973). In other words, increasing oxygen content progressively stabilizes $\beta$ against the lattice collapse associated with the $\beta \rightarrow \omega$ transformation pathway. This parallels trends in Ti–16Nb alloys, where moderate oxygen additions initially suppress $\omega$ formation but allow its reappearance once the martensitic transformation is inhibited, implying a shift in the transformation pathway towards $\beta$ decomposition into $\omega$ rather than $\beta \rightarrow \alpha''$.

Overall, the combined XRD results demonstrate that oxygen strongly modifies transformation pathways in Ti–Nb alloys, revealing the nonlinear effect of oxygen in suppressing $\alpha''/\omega$ formation. Diffraction analysis reveals suppression or reappearance of $\omega$ and $\alpha''$ phases depending on composition, while SEM observations confirm the corresponding morphological manifestations.

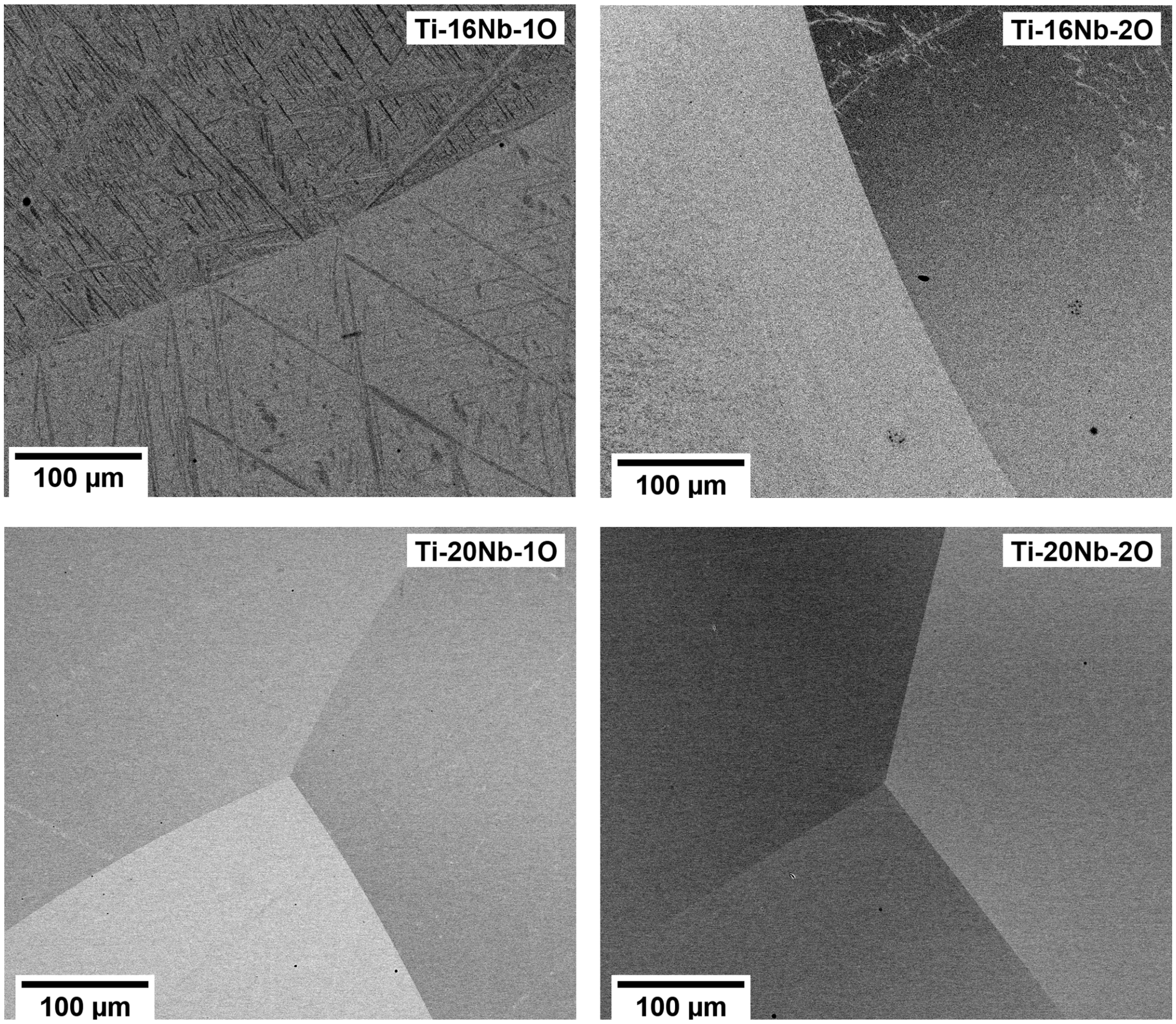

**Figure 11**
BSE images of Ti–(16–20)Nb–*x*O samples after solution treatment and water quenching.

### 4.3. Quantitative analyses – indexing

To distinguish individual variants of martensite from one another, we simulated reciprocal space in the range from $(hkl) = (\overline{7}\,\overline{7}\,\overline{7})_{\alpha''}$ to $(hkl) = (777)_{\alpha''}$. All the zero intensity reflections due to extinction rules were neglected at the very beginning to speed up the calculations. The way of determining the crystal orientation was described above in Section 3.1; here we show the results and procedure on a Ti–20Nb sample.

The initial step involved identifying the $2\theta$ and $\xi$ angles of all measured peaks. To accomplish this, we employed the *Find Peaks* function, applying threshold values for both peak intensity and prominence. As observed (Fig. 12), not all peaks were detected by the algorithm, primarily due to the relatively high intensity threshold set to suppress unwanted peaks. Nevertheless, this limitation did not pose a significant issue, as a sufficiently large number of other peaks remained available for comparison with the simulated data.

In the next step, we simulated the experimental reciprocal-space map of the 12 mutually oriented variants of martensite. Because we did not know the initial orientation of the prior $\beta$ grain, our algorithm simulated many diffraction patterns which would arise during rotation of the sample within [−30°, 30°] around the horizontal axis. This was done several thousand times (~20000) with different initial orientations of the parent $\beta$ crystal. The result with the highest-scoring parameter (number of matching pairs) is shown in Fig. 13.

We can see from the resulting simulated pattern that most points were matched with a precision of several pixels and the predicted positions have a good agreement with the measured reciprocal-space map. Even though some points can be found which are present in the map but not predicted by the simulation, it could be the case that these spots were generated near the limits of the [−30°, 30°] range of rotation and thus, due to the precision of our measurements, they do not have to be included in our simulation. The other limits come from errors within the experimental arrangement – manual adjustment of the eccentric position of the sample, or the precision of the sample preparation, which due to its low thickness could be easily bent and thus affect the widths and positions of the peaks. Since it is hard to quantify such errors, we decided not to index such peaks as this indexing might be ambiguous. The resulting part of the 2D image plate with simulated positions of reflections is presented in Fig. 14.

This is how we managed to index the reciprocal-space map in the case of samples which did not contain any phase other than $\alpha''$. In the case of samples containing $\beta$ or $\omega$ phases, we could not always successfully index the reciprocal-space maps due to large errors caused by the presence of extraneous peaks. As a result, the possibilities of determining the crystal structure were narrowed down to only six samples, namely Ti–20Nb, Ti–12Nb–0/1/2/3O and Ti–16Nb–1O.

### 4.4. Quantitative analysis – elucidation of $y$ parameter as $\{110\}_\beta$ shuffle

Following the successful indexing of the reciprocal-space maps via peak positions, the analysis focused on quantifying the atomic $y$ coordinate, which requires determination of structure factors from peak intensities. Integral intensities of individual diffraction maxima were extracted to determine absolute experimental structure factors as described in Section 3.2. Each reciprocal-lattice point was modelled using a function consisting of a planar background and a 2D elliptic Gaussian with a rotation parameter relative to the horizontal axis, allowing accurate representation of the observed peak morphology and proper background subtraction (Fig. 15).

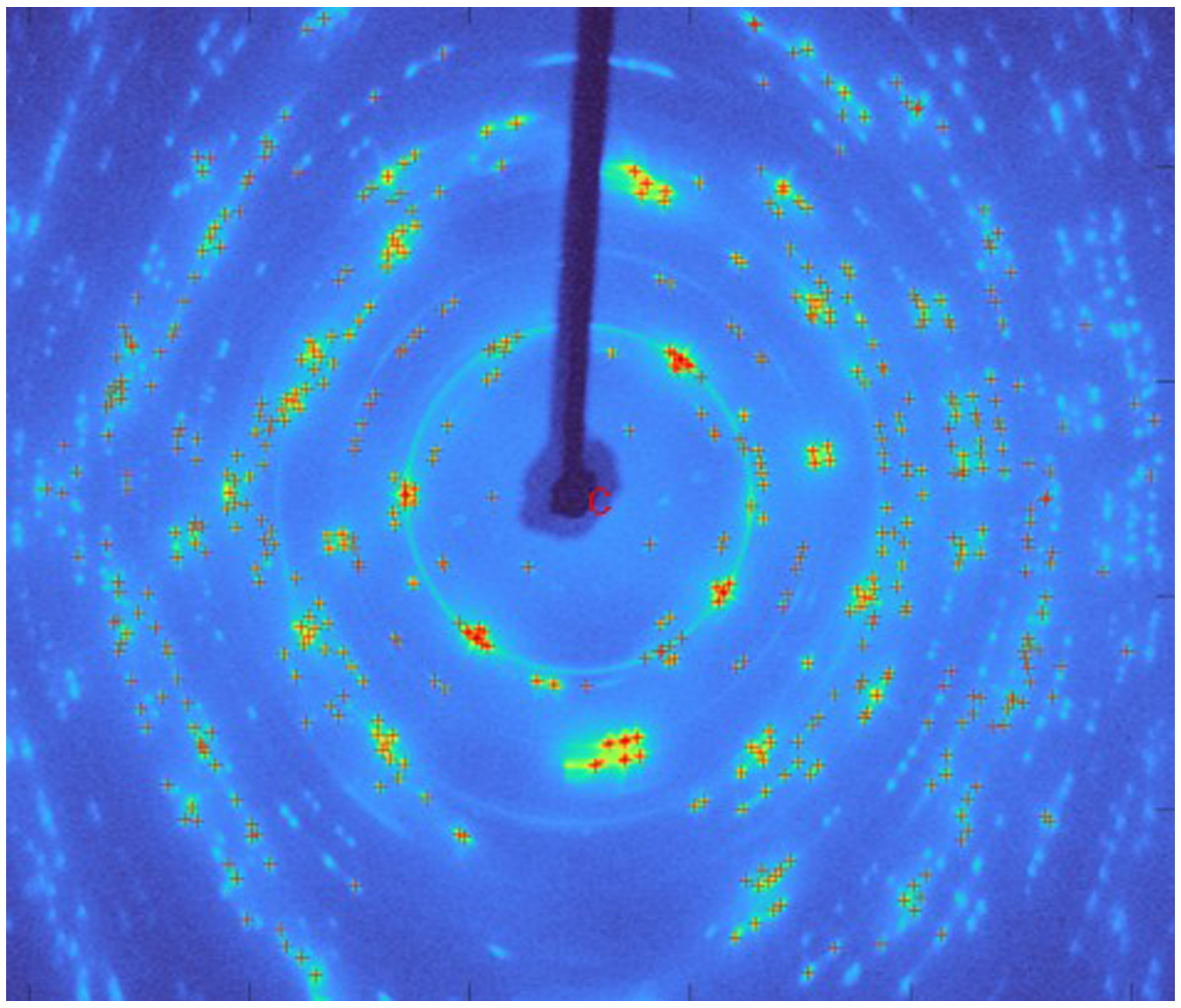

**Figure 12**
Measured signal on a 2D XRD detector. Deviation in the horizontal range is equal to the typical $2\theta$ angle. Red crosses represent the found locations of peaks.

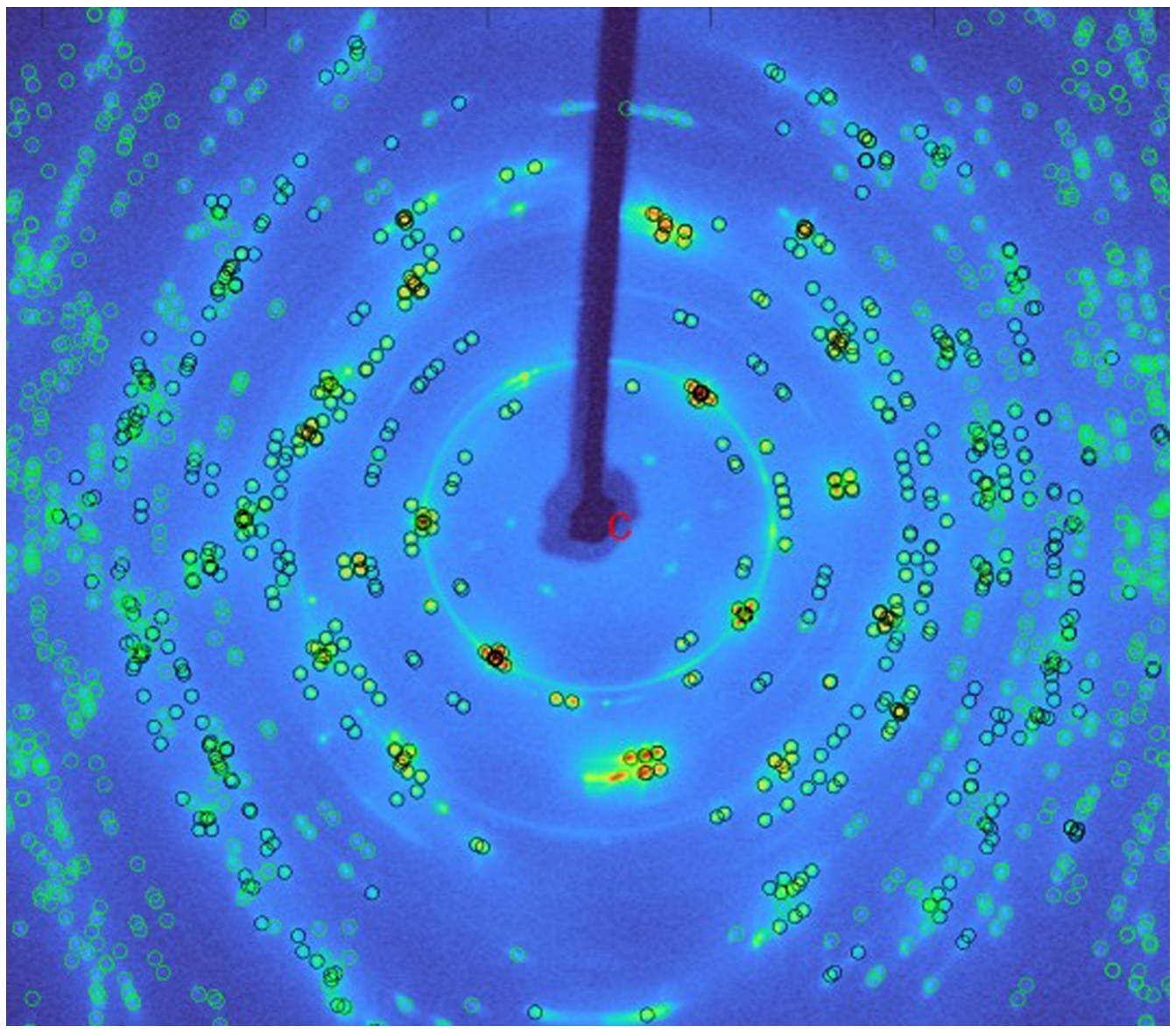

**Figure 13**
XRD measurement overlaid by black circles (matched pairs of reciprocal points generated by simulation) and green circles (predicted positions of diffraction peaks which should be measured within [−30°, 30°] rotation).

**Table 3**
Comparison between observed (obs) and calculated (calc) structure factors for the Ti–20Nb sample, with standard deviations calculated using equivalent reciprocal-space points; $y$ = 0.212.

| $\overline{\lvert F_{hkl}^{\rm obs}\rvert}$ | $\lvert F_{hkl}^{\rm calc}\rvert$ | $\sigma_{\overline{\lvert F_{hkl}^{\rm obs}\rvert}}$ | $h$ | $k$ | $l$ | $\lvert\overline{\lvert F_{hkl}^{\rm obs}\rvert} - \lvert F_{hkl}^{\rm calc}\rvert\rvert$ |
|---|---|---|---|---|---|---|
| 16.69 | 16.50 | 0.84 | 1 | 1 | 0 | 0.20 |
| 26.82 | 26.82 | 1.55 | 0 | 2 | 1 | <0.01 |
| 9.52 | 10.89 | 1.26 | 1 | 1 | 2 | 1.37 |
| 20.76 | 21.87 | 1.81 | 1 | 3 | 0 | 1.11 |
| 20.07 | 20.60 | 0.52 | 1 | 3 | 1 | 0.53 |
| 24.32 | 23.13 | 2.70 | 2 | 2 | 0 | 1.19 |
| 12.28 | 12.70 | 2.97 | 0 | 2 | 3 | 0.41 |
| 10.81 | 12.00 | 1.55 | 2 | 2 | 1 | 1.19 |
| 15.72 | 15.51 | 1.96 | 1 | 3 | 2 | 0.21 |
| 11.23 | 11.23 | – | 0 | 4 | 0 | <0.01 |
| 17.17 | 17.17 | 3.38 | 0 | 4 | 1 | <0.01 |
| 4.32 | 3.90 | – | 1 | 1 | 4 | 0.42 |
| 3.54 | 3.52 | – | 3 | 1 | 0 | 0.02 |
| 5.37 | 6.35 | 0.71 | 2 | 2 | 3 | 0.98 |

**Table 4**
Refined parameters from XRD for the experimental alloys Ti–12Nb–$x$O, Ti–16Nb–1O and Ti–20Nb.

| Sample | $\overline{y}$ | $\sigma_y$ | $\eta_S$ |
|---|---|---|---|
| Ti–12Nb–0O | 0.189 | 0.006 | 0.738 |
| Ti–12Nb–1O | 0.184 | 0.003 | 0.797 |
| Ti–12Nb–2O | 0.189 | 0.008 | 0.726 |
| Ti–12Nb–3O | 0.185 | 0.002 | 0.785 |
| Ti–16Nb–1O | 0.197 | 0.006 | 0.640 |
| Ti–20Nb–0O | 0.212 | 0.005 | 0.461 |

The integral intensity was then obtained by integrating the Gaussian component after refinement of all peak parameters. To preserve data integrity, only reflections sufficiently isolated from neighbouring predicted spots were included, to prevent intensity overlap. Fit quality was evaluated for each peak and reflections with large standard deviations were excluded from subsequent structural refinement.

To eliminate the assumption of a homogeneous volume fraction across all 12 martensitic variants, the structural parameters were refined independently for each variant. This approach accounts for non-random variant selection effects and texture, both potentially arising during the $\beta\rightarrow\alpha''$ transformation.

A limitation of the variant-specific refinement is the reduced number of non-overlapping reflections per variant, resulting in fewer integral intensities. The representative value of the $y$ coordinate was obtained by averaging the values refined for the individual variants. The refined theoretical values of the structure factors are listed in Table 3.

Beyond the expansion and contraction of the lattice vectors, the transition from hexagonal to orthorhombic symmetry is fundamentally governed by the internal rearrangement of atoms within the unit cell, as presentented in Section 2. During the refinement of the $\alpha''$ phase (space group $Cmcm$), the $y$ coordinate of the Wyckoff position 4$c$ was treated as a free parameter to account for this atomic shuffle. In a structurally ideal hexagonal lattice, this position is fixed at $y_{\alpha'} = \frac{1}{6} = 0.166$, but our refinements reveal a consistent increase in $y$ as the Nb concentration rises. As detailed in Table 4 and plotted in Fig. 16, the calculated $y$ values for the Ti–Nb series exceed the hexagonal limit, moving from approximately 0.189 in Ti–12Nb towards 0.212 in the more concentrated Ti–20Nb alloy, almost reaching the $\beta$ limit of $y_\beta = \frac{1}{4} = 0.250$.

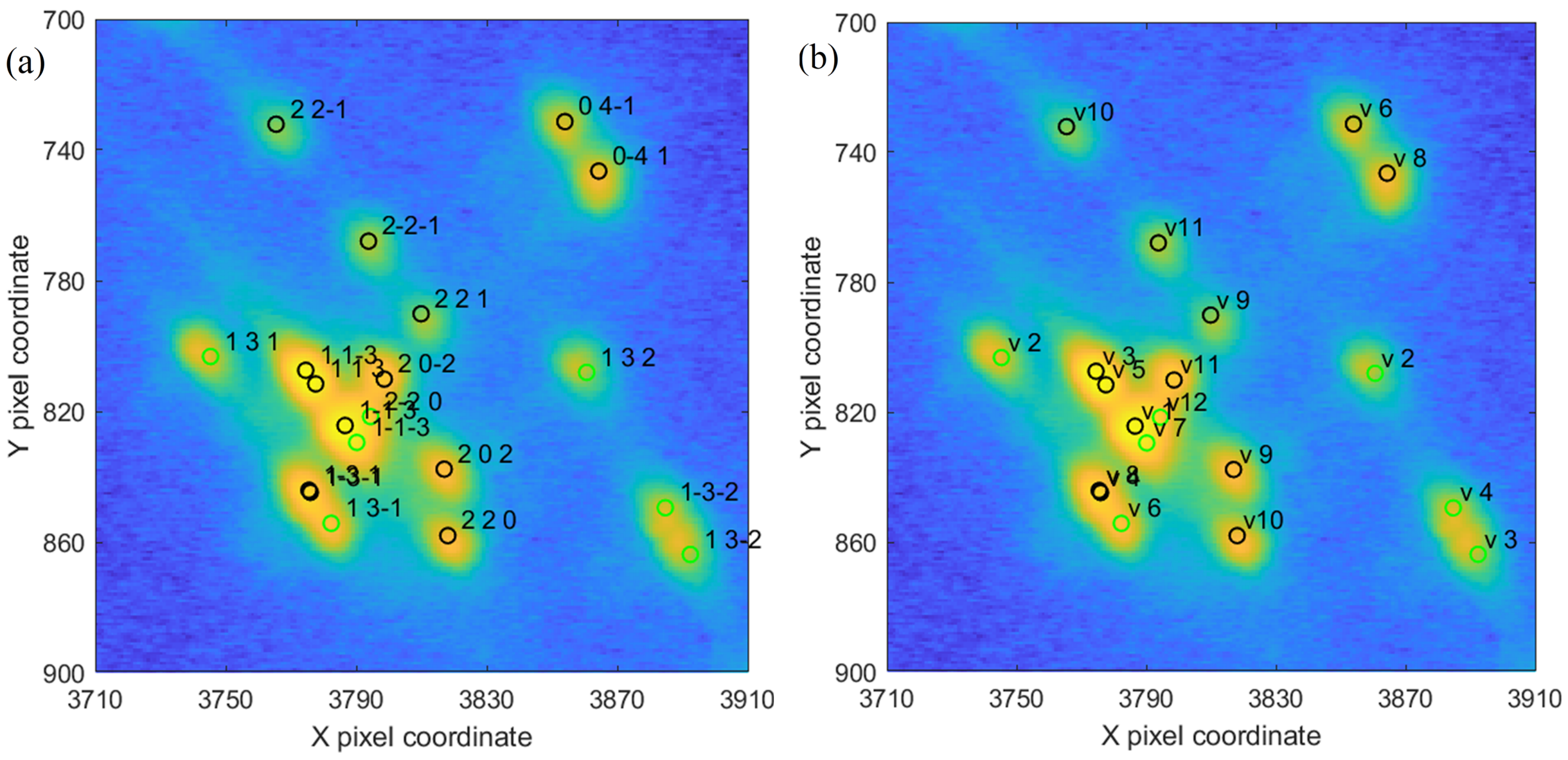

**Figure 14**
Part of the 2D image plate overlapped with the simulated pattern showing resolved reflection planes ($a$) belonging to different variants ($b$).

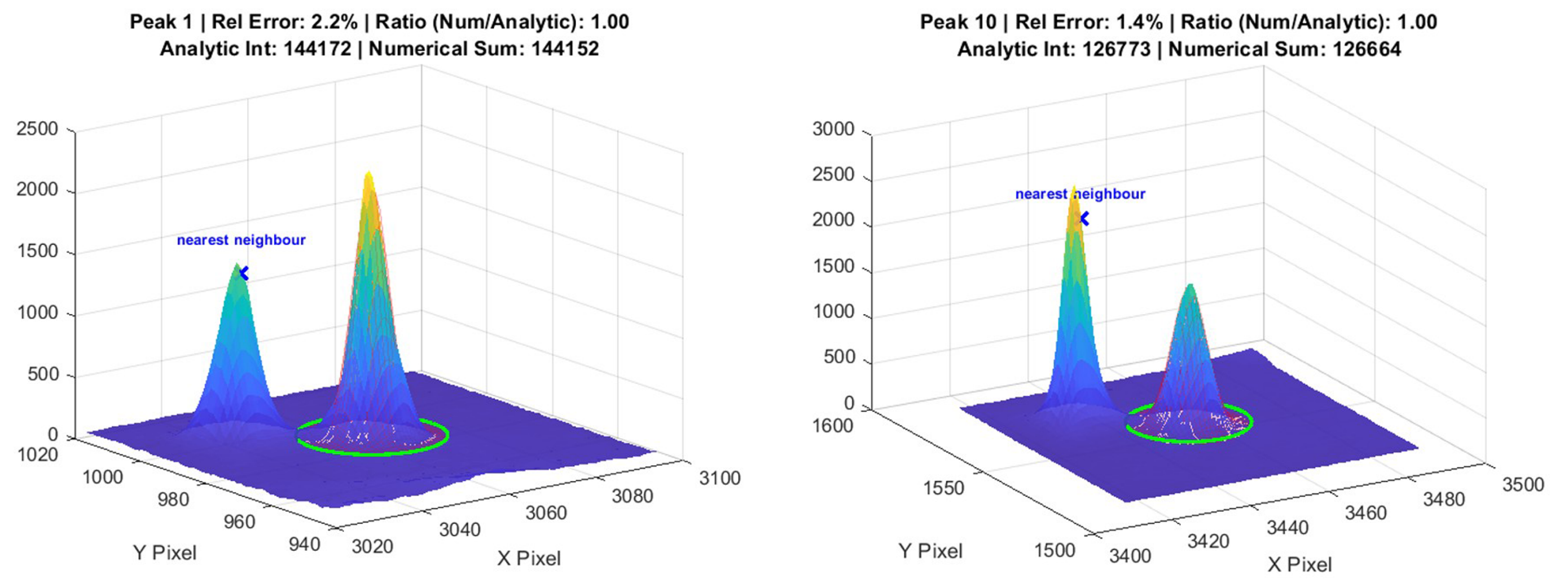

**Figure 15**
Isolation and Gaussian fitting of peak intensities within the encircled green area, excluding nearest-neighbour overlap, to determine the integral intensity. Relative error and the ratio of numerical to analytical integration were evaluated, ensuring data integrity.

This trend indicates that higher solute concentrations hinder the atoms from reaching the positions required for hexagonal symmetry during the $\beta$ to $\alpha'$ transformation, manifested as a decrease in the orthorhombic parameter $\eta_S$ from about 0.74 for Ti–12Nb to 0.46 in Ti–20Nb. Physically, this shift is driven by the interatomic potential of the surrounding Nb and Ti atoms, as the Nb content increases the tendency of the lattice to retain the eightfold coordination of the parent $\beta$ phase. This results in a larger atomic displacement along the $b$ axis, effectively 'locking' the structure into an orthorhombic configuration and preventing the completion of the shuffle towards the $\alpha'$ hexagonal state. This internal structural lag correlates directly with our observed decrease in the $b/a$ ratio, confirming that the $y$ parameter is a parallel descriptor of the degree of orthorhombicity in these martensitic alloys. These results are consistent with a previous study performed on Ti–Nb alloys (Bönisch *et al.*, 2014).

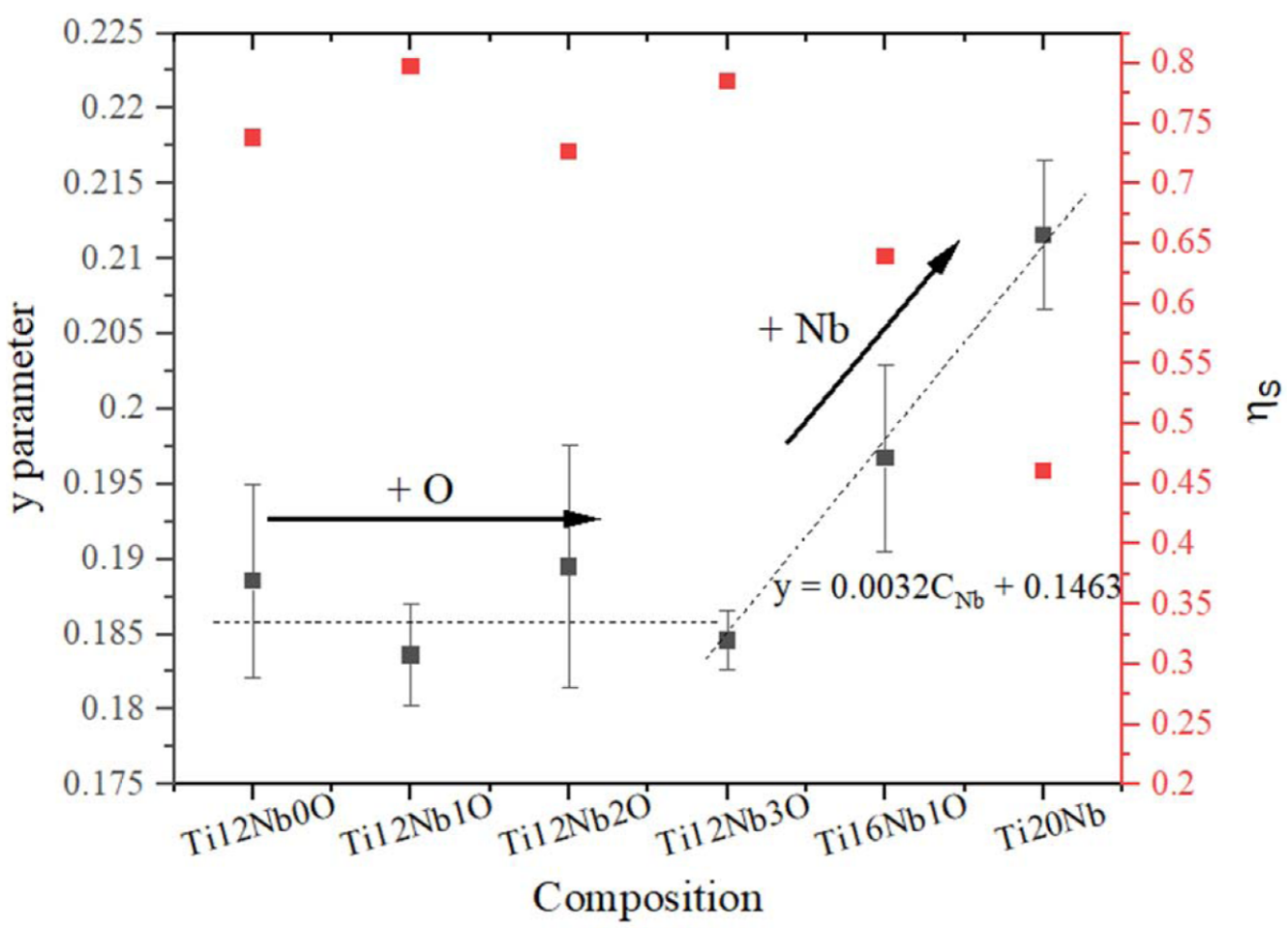

**Figure 16**
Refined parameters from XRD for the experimental alloys Ti–12Nb–$x$O, Ti–16Nb–1O and Ti–20Nb.

In contrast, the Ti–12Nb–$x$O alloys exhibit no systematic dependence of the $y$ parameter on the oxygen content within experimental uncertainty. The refined values remain constant across the investigated oxygen concentrations and any potential changes fall within the refinement error. This indicates that oxygen addition does not produce a measurable modification of the atomic shuffle along the $b$ axis under the present experimental conditions.

## 5. Conclusions

This work has investigated the influence of niobium and oxygen on phase stability and crystal structure in Ti–Nb–O alloys subjected to solution treatment and quenching. Fifteen alloys with compositions spanning Ti–(8–28)Nb–(0–3)O at.% were characterized primarily by X-ray diffraction, complemented by microscopy and crystallographic analysis where applicable.

The results demonstrate that the lattice of orthorhombic $\alpha''$-martensite evolves continuously with decreasing niobium concentration, bridging the crystallographic characteristics of the parent b.c.c. $\beta$ phase and the stable $\alpha'$ h.c.p. structure. This behaviour confirms the intermediate nature of $\alpha''$ within the transformation sequence between the two equilibrium phases.

Systematic variation of the oxygen concentration revealed a general suppressing effect on both $\beta\rightarrow\omega$ and $\beta\rightarrow\alpha''$ transformations, although the dominant pathway depends on the niobium content. At lower niobium content ($C_{Nb} \lesssim 12\%$), oxygen primarily acts to suppress the $\beta\rightarrow\omega$ transformation and redirects the structure toward the pure martensitic $\alpha''$ phase, whereas in more $\beta$-stabilized alloys with high niobium content ($C_{Nb} \gtrsim 20\%$), oxygen inhibits the long-range growth of macroscopic $\alpha''$. Because this primary transformation pathway is blocked, the unstable $\beta$ matrix instead partially decomposes into the $\omega_{ath}$ phase.

These observations are consistent with the role of oxygen as an interstitial defect generating local stress fields that hinder long-range martensitic transformation rather than producing a significant change in the intrinsic $\beta$ or $\alpha''$ crystal structure. Accordingly, no measurable change in atomic positions was detected with increasing oxygen content.

In contrast, niobium content exerted a measurable and continuous effect on the crystallographic parameters and atomic shuffle associated with the $\{110\}_\beta\langle 110\rangle_\beta$ mechanism. The shuffle parameter increased with Nb concentration, reflecting structural evolution toward $\beta$-phase symmetry and highlighting the strong coupling between shuffle and shear processes governing transformation crystallography.

Overall, this combined structural and microstructural analysis establishes that niobium primarily controls continuous lattice evolution and martensitic crystallography, whereas oxygen acts as a defect-mediated stabilizer that suppresses competing transformations without significantly altering the intrinsic crystal structure of $\alpha''$-martensite. These findings contribute to a clearer understanding of transformation mechanisms in metastable Ti alloys and provide guidance for tailoring phase stability through interstitial and substitutional alloying.

## Acknowledgements

During the preparation of this work the authors used ChatGPT to correct the grammar and polish the sentences. After using this tool/service, the authors reviewed and edited the content as needed, and they take full responsibility for the content of the published article. Open access publishing facilitated by Univerzita Karlova, as part of the Wiley–CzechELib agreement.

## Data availability

The data that support the findings of this study are openly available in the Zenodo repository at https://doi.org/10.5281/zenodo.19087995 under a CC-BY 4.0 licence.

## Funding information

This work was supported by the Czech Science Foundation (project No. 21-18652M). Additional financial support was provided by the Operational Programme Johannes Amos Comenius of the Ministry of Education, Youth and Sports of the Czech Republic, within the project Ferroic Multifunctionalities (FerrMion) (project No. CZ.02.01.01/00/22_008/0004591), co-funded by the European Union. Kristián Šalata acknowledges support from the Grant Agency of Charles University (project No. 282122). The authors also acknowledge the use of the MGML facilities (https://mgml.eu/), supported within the programme of Czech Research Infrastructures (project No. LM2023065).